\documentclass[prl,showpacs,twocolumn, superscriptaddress]{revtex4}  
\usepackage{graphics}
\usepackage{amsmath}

\begin{document}

\title{A multibeam atom laser: coherent atom beam splitting from a single far detuned laser }

\author{J. Dugu\'e}
\affiliation{Australian Research Council Centre Of Excellence for Quantum-Atom Optics, Department of Physics, The Australian National University, Canberra, ACT 0200, Australia}
\affiliation{Laboratoire Kastler Brossel, ENS, UPMC-Paris 6, CNRS, 24 rue Lhomond, 75231 Paris Cedex 05, France}

\author{G. Dennis}
\author{M. Jeppesen}
\author{M. T. Johnsson}
\author{C. Figl}
\email{cristina.figl@anu.edu.au}
\homepage{http://www.acqao.org}
\author{N. P. Robins}
\author{J. D. Close}
\affiliation{Australian Research Council Centre Of Excellence for Quantum-Atom Optics, Department of Physics, The Australian National University, Canberra, ACT 0200, Australia}

\begin{abstract}
We report the experimental realisation of a multibeam atom laser.
A single continuous atom laser is outcoupled from a Bose-Einstein condensate (BEC) via an optical Raman transition.
The atom laser is subsequently split into up to five atomic beams with slightly different momenta, resulting in multiple, nearly co-propagating, coherent beams which could be of use in interferometric experiments.
The splitting process itself is 
a novel realization of Bragg diffraction, driven by each of the optical Raman laser beams independently.
This presents a significantly simpler implementation of an atomic beam splitter, one of the main elements of coherent atom optics.
\end{abstract}


\maketitle
Diffraction of atoms from an optical grating has led to a wealth of insights into atomic physics, and to practical applications such as coherent beam splitting for precision atom interferometry~\cite{kasevich1,sackett,giltner2}. 
Bragg diffraction of thermal atomic beams has been done mostly under nearly normal incidence to the optical grating~\cite{pritchard, oberthaler, oberthaler2,giltner}.
In these experiments, great care had to be taken to meet the Bragg condition by precisely matching the angle of the optical grating to the velocity of the atoms. 
With the advent of dilute gas BEC~\cite{cornell,ketterle}, a source of cold atoms emerged that is ideal for studying and utilizing atom/light interactions in a highly controllable way. 
Investigations of BEC diffraction with light centered on diffraction at normal incidence using short light pulses, leading, among other results, to the observation of periodic focusing, collimation and the atomic Talbot effect~\cite{phillips1,phillips2}.
It was soon realised that controllable Bragg diffraction could be generated in a stationary condensate by applying an optical running wave~\cite{phillips3}.
Energy and momentum conservation for diffraction are met by precisely setting the energy (frequency) difference between the two incident lasers.  This is equivalent to satisfying the Bragg condition on the angle of incidence in atomic beam experiments. 
Bragg and Raman diffraction from two detuned optical beams became an immensely powerful tool.  It has been used as a coherent beam splitter in interferometry~\cite{kasevich1,sackett,giltner2}, as a spectroscopic probe of a BEC~\cite{ketterle5,phaseFlucAspect}, as the basis for demonstrating superradiance and matter wave amplification~\cite{RamanAmplification, ketterle4, ketterle3, Kozuma}, as a tool to measure the relative phase between two BECs~\cite{ketterle2}, and as a mechanism for producing~\cite{phillipsRaman} and manipulating~\cite{opticsAL,braggALtheory} an atom laser.   

\begin{figure}[b]
\centerline{\scalebox{.4}{\includegraphics{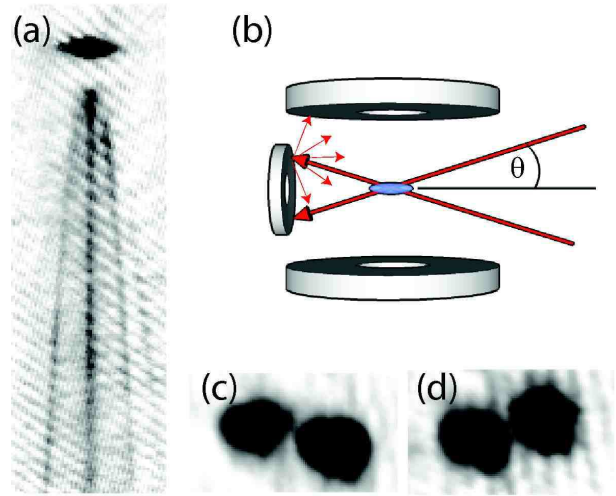}}}
\caption{(a): Absorption image (360\,$\mu$m x 1100\,$\mu$m) of a multibeam atom laser derived from 20ms continuous Raman output coupling (b): Experimental setup showing optical lasers, BEC (not to scale), and magnetic coils (c), (d): Absorption image of a pulsed RF atom laser, exposed to the upward (c) or downward (d) propagating Raman laser only. Image size is 540\,$\mu$m x 320\,$\mu$m, and $\theta=15^\circ$}
\label{fig1}
\end{figure}

In this article we present a mechanism for highly efficient cw Bragg diffraction of a continuous atom laser under non-normal incidence onto an optical grating.
As opposed to previous realisations of Bragg diffraction, the gratings are each produced from only a single far detuned optical laser and its diffuse backreflection.
In Figure 1(a) we show an absorption image of the system in operation.
An atom laser beam outcoupled from a $^{87}$Rb BEC is seen in the centre and the diffracted beams off to the sides, falling under gravity.
We outcouple the initial atom laser using two optical beams that drive a two-photon Raman transition~\cite{phillipsRaman,nick}, coherently transfering atoms from the magnetically trapped $|F=1,m_F=-1\rangle$ to the magnetically untrapped $|F=1,m_F=0\rangle$ state, forming a continuous atom laser beam~\cite{esslinger, aspect,nick2}.
The laser beams are roughly collimated and have a radius of around 500\,$\mu$m, much bigger than the BEC.
The splitting, as opposed to the output coupling, is induced by the subsequent interaction of the atoms with the laser beams individually:
the upward (downward) propagating laser imparts a momentum into (against) the laser direction resulting in the atom beam on the left (right) as illustrated in Figure 1 (c) and (d) where we illuminated an rf-outcoupled atom laser pulse with only one of the laser beams. 
After 20\,ms of free fall an absorption image of the pulse reveals two momentum components.

The source for the atom laser is a BEC of $5 \times 10^5$~atoms.
We use a highly stable, water cooled QUIC magnetic trap (axial frequency $\omega_{y} = 2 \pi \times 13$~Hz and radial frequency $\omega_{\rho} = 2 \pi \times 130$~Hz, with a bias field of $B_0=2$~G).
The orientations of our magnetic coils and the optical Raman beams are shown in Figure~\ref{fig1} (b).
The two laser beams propagate under an angle $\theta$ to the horizontal in the plane of gravity and the magnetic trap bias field. 
They are produced by a 700~mW diode laser red-detuned by 300 GHz from the D2 transition in $^{87}$Rb.
We turn the laser power on or off in less than 200~ns using a fast switching acousto-optic modulator (AOM) in a double pass configuration.
After the switching AOM, the light is split and sent through two separate AOMs, each again in a double pass configuration.
The frequency difference between the AOMs corresponds to the Zeeman plus kinetic energy difference between the initial and final states of the two-photon Raman transition.
We stabilize the frequency difference by  running the 80~MHz function generators driving the AOMs from a single oscillator.
The beams are then coupled via single mode, polarization maintaining optical fibers directly to the BEC through a collimating lens and waveplate, providing a maximum intensity of 2500~mW/cm$^2$ per beam at the BEC.
The polarization of the beams is optimized to achieve maximum outcoupling,  maximizing $\pi$ polarization for the downward propagating beam and $\sigma^-$ for the upward propagating beam. 
Continuously applying the Raman output coupler to the condensate produces the image in Figure~\ref{fig1}(a);
a continuous atom laser coherently split into three (or more) co-propagating momentum states. As we will show, the splitting is caused by Bragg diffraction from two optical standing wave gratings.

In our setup, we do not align a second laser beam to set up a grating for the Bragg diffraction.
Also, we can rule out a direct back reflection of the beam from the geometry of the set up, and
a careful measurement searching for a direct back reflection of the input beam has not revealed any measurable effect.
However, the beams hit the magnetic coils, giving rise to diffuse backscattering. 
For a small range of angles, the beam can pass through the atoms without clipping on the apparatus.
In such a situation, we observe no measurable transfer into the momentum side-mode.
But deliberately placing black cardboard as a diffuse scatterer in the path of the laser, after it has passed through the glass cell, brings the diffraction back.
We estimate that in our experimental setup, depending on $\theta$, determining the distance of the scatterer to the interaction region, the backscattered intensity to be between 0.01\% and 0.06\% of the incoming intensity.

It might seem surprising at first that such a small fraction of reflected light induces an efficient transfer into higher order momentum modes.
From Raman outcoupling experiments, where both lasers are applied to the condensate, we have calibrated the 2-photon Rabi frequency, $\Omega_{\rm 2P} = \frac{\Omega_1 \Omega_2}{\Delta}$, where $\Omega_{1,2}$ are the one-photon Rabi frequencies. With maximum laser intensity we can achieve a maximum of $\Omega_{\rm 2P} = 2\pi\times 40$\, kHz. 
Thus, we estimate the combination of the incident laser and the diffuse backscattered light to be able to drive a maximum two photon Rabi frequency of $\Omega \approx 2\pi\times $1\,kHz (because the single-photon Rabi frequency scales with the square root of the light intensity).
In a naive model, the atoms are assumed to Rabi-flop between the two momentum states.
We will later point out the limits of this model and introduce a more correct one.
The time-dependent population of a Rabi-flopping state is
$\sin^2(\Omega t)$.
For a 1\,ms laser pulse, this results in a diffraction efficiency of more than 70\%.
The estimate makes it plausible that the seemingly small amount of backscattered light gives rise to efficient coupling.

There are a few characteristic properties of Bragg diffraction that we can expect in the experiment.
Diffraction from a standing wave grating occurs from absorption and emission of a photon, leading to a momentum transfer to the atom of $2\hbar \bf{k}$ . 
The kinetic energy of the atom, however, must remain unchanged since the energy of the absorbed and the emitted photon are the same and the atoms stays in the same internal state.
This condition is only fullfilled when the atom is traveling with the resonance velocity 
\begin{equation}
	v_{\rm res} = \pm \frac{\hbar k/m}{\sin{(\theta})},
	\label{eq.vres}
\end{equation}
where $m$ is the atomic mass, and ($\pi /2 - \theta$) is the angle of the laser beam with the velocity of the atom. This is when Bragg diffraction can occur.

\begin{figure}
\centerline{\scalebox{.45}{\includegraphics{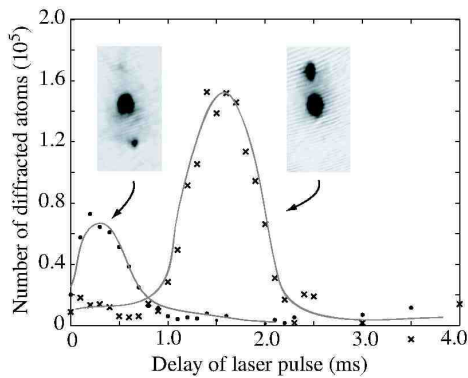}}}
\caption{Number of diffracted atoms  as a function of pulse delay relative to trap switch off. Visible are the two possible velocity resonances, for atoms traveling upwards (circles) and then traveling downwards (crosses). Pulse duration is 300$\mu s$ (the delay is given from the start of the pulse), and $\theta=70^o$. 
The insets show the corresponding images, each 460\,$\mu$m x 890\,$\mu$m. The lines are interpolants to guide the eye.}
\label{fig.delay}
\end{figure}

In order to investigate the mechanism responsible for the coherent splitting and to test it against the Bragg characteristics, we separated the outcoupling mechanism from the splitting. 
We release the complete BEC by switching off the magnetic trap.
Upon trap switch-off, the atoms are accelerated upwards, resulting in a launch velocity of 1.1\,cm/s~\cite{switchoffkick}.
Thus, for $\theta>35^o$, the atoms go through two velocity resonances: 
the first diffraction occurs when the atoms are traveling upwards, the second when they travel downwards.
Since the resonance velocity increases with decreasing $\theta$, for $\theta<35^o$ the atoms move upwards too slowly to be diffracted.
After a variable delay time we pulse only one of the Raman lasers on.
We then allow the atoms with the different momentum components to separate in free fall before taking an absorption image 22\,ms after the magnetic trap switch-off.  
From the images, we calculate the total number of transmitted and diffracted atoms.

First, we vary the pulse delay while keeping the laser power and pulse duration fixed, for a few different angles.
The results of one of these measurements, for $\theta=70^o$, are shown in the main plot of Figure~\ref{fig.delay}:
We observe two resonance structures at delay times which match the resonance velocities.
The width of the resonance is consistent with the energy spread of the trapped BEC.
For all measurements, the momentum transfer to the diffracted atoms as well as the timing of the resonance is consistent with Bragg diffraction.

\begin{figure}
\centerline{\scalebox{.45}{\includegraphics{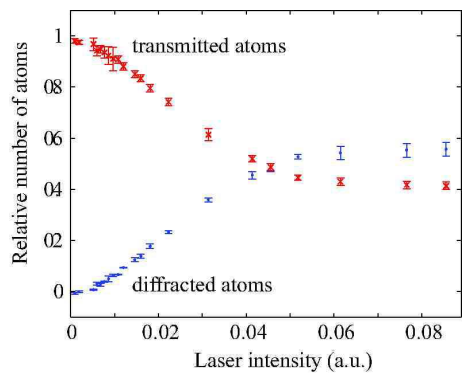}}}
\caption{Diffraction efficiency as a function of laser intensity for fixed pulse duration of 2ms and  $\theta=15$.}
\label{fig.intensity}
\end{figure}

Second, fixing the delay time, angle and pulse duration, we measure the diffraction efficiency as a function of the laser intensity (Figure~\ref{fig.intensity}).
For a 2\,ms laser pulse with the delay to the trap switch off chosen such as to address the resonance velocity, and $\theta = 15^o$, we measure a transfer efficiency of up to 60\% into the momentum side-mode, consistent with our simple estimate of diffraction efficiency above.

\begin{figure}
\centerline{\scalebox{.38}{\includegraphics{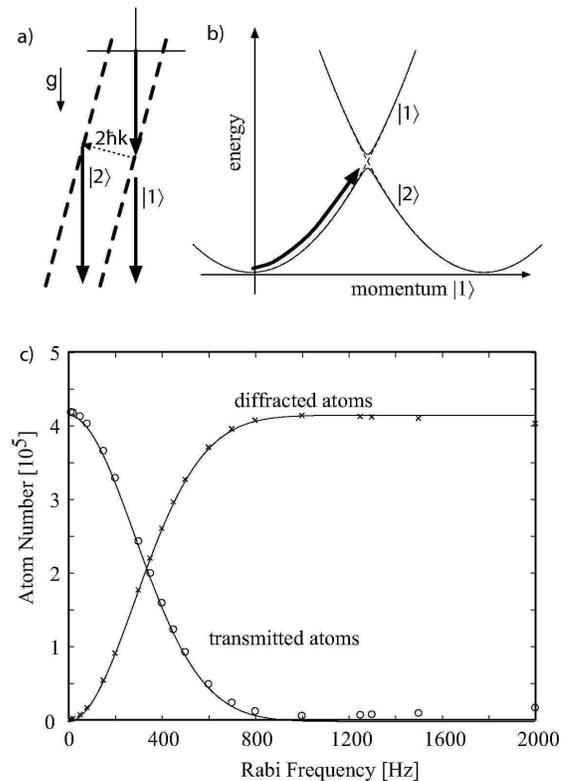}}}
\caption{a) Resonances in momentum space. Diffraction can occur where the atom momentum trajectory intersects with the dashed line. b) Avoided crossing for $\theta = 90^o$. c)
GP simulation results (symbols). Solid lines are the  predicted Landau-Zener transition probabilities. $\theta = 15^o$}
\label{fig.theory}
\end{figure}

This simple picture, however, does not take into account the velocity selectivity of the process. A more accurate description of the process results in a situation equivalent to an avoided crossing between the diffracted and undiffracted atoms. If we consider the transfer of atoms during the diffraction process, the diffracted atoms can be considered to be in a distinct state from the undiffracted atoms because their momentum difference, $2 \hbar \bf{k}$, is significantly greater than the momentum width of the falling atoms, see Figure~\ref{fig.theory}(a) for an illustration of the resonances in this system. The diffracted ($|1\rangle$) and undiffracted ($|2\rangle$) atom states are coupled by the diffraction grating leading to an avoided crossing in momentum space, as shown in Figure~\ref{fig.theory}(b). As the atoms are in free fall, their momentum, and hence the energy difference between the two levels will vary linearly in time, enabling the Landau-Zener theory~\cite{landau,zener} to be applied. In this situation, the Landau-Zener theory gives the diffraction probability as
\begin{equation}
P = 1- \exp\left( - \frac{\pi \Omega^2}{\left | \mathbf{k} \cdot \mathbf{g} \right |}\right),
\end{equation}
where $\Omega$ is the two-photon Rabi frequency, $\mathbf{k}$ is the wave vector of the photons and $\mathbf{g}$ is the acceleration due to gravity. Hence, for a sufficiently high two-photon Rabi frequency, this model predicts a perfect transfer of atoms from the undiffracted state to the diffracted state.
We do not consider a second diffraction of the diffracted pulse or the diffraction into higher orders because they occur at different resonance velocities which are not reached while the optical grating is present.

We have verified the validity of the Landau-Zener model by solving the Gross-Pitaevskii equation (in two dimensions) for the diffracted and undiffracted atomic states, including the effect of s-wave scattering between the atoms. The results from both approaches agree very well and are shown in Figure~\ref{fig.theory} (c). For high Rabi frequencies, it can be seen that the transition probability decreases from the result from Landau-Zener theory due to power broadening causing the state changing to begin closer to the BEC. Here, the inter-particle interactions are non-negligible. 
The theoretical results reproduce the general shape of the experimental data, in particular the plateau towards high Rabi frequencies. However, there is some discrepancy in the maximum transfer probability which we attribute to the fact that the diffuse nature of the reflection is not taken into account in the calculations.
The simulations confirm the estimate of the reflected light intensity and the resulting Rabi frequency.

\begin{figure}
\centerline{\scalebox{.25}{\includegraphics{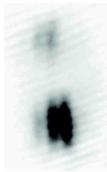}}}
\caption{Spatial modulations occuring for large $\theta$ and high light intensities. Size of the image is 480\,$\mu$m x 620\,$\mu$m.}
\label{fig.structure}
\end{figure}

We would like to mention one observation that we cannot explain at this point. 
We observe the appearance of strong density modulation of both the condensate and the diffracted pulse when the angle $\theta$ and the light intensity become large as shown in Figure~\ref{fig.structure}.
The wavelength of the modulation is several 10\,$\mu$m.
Interestingly, the structure of the diffracted pulse is out of phase with the structure on the BEC.
At this point, we do not know if the spatial structure is related at all to the splitting mechanism and we consider taking more data to illuminate the nature of the underlying physics.

To the best of our knowledge, high efficiency atomic diffraction from a single laser and its own diffuse backscattering has not been previously observed.
The method presents a novel, experimentally simple and versatile tool for atom optics, and we have used it to produce a multibeam atom laser.
Compared to previous atomic beam splitters which are based on two (usually counter-propagating) optical laser beams, our method presents a significant experimental simplification.
In vacuum systems which do not allow optical access from two  sides it may be the only possibility to implement an atomic beam splitter.
The direction of the momentum transfer can be controlled, together with the resonance velocity and thus the resonance position, by the angle of the incoming laser beam.
Future applications can be envisaged in coherent atom interferometry with a separated paths interferometer.
For this purpose, a Raman outcoupled atom laser beam is perfectely suited, e.g. offering a larger brightness, and a good beam profile~\cite{nick}. 
Our results show that it is possible to run the output coupling laser beams in a regime where they serve at the same time as two beam splitters thus significantly simplifying the experimental demands. 

This work was supported by the APAC National Supercomputing Facility and by the Alexander von Humboldt Foundation. We acknowledge very fruitful discussions with Peter Drummond.


\begin{thebibliography}{99}
\bibitem{switchoffkick}We determined the launching velocity by measuring the difference in position of a pulse of atoms in the $m_F=1$ and the magnetically insensitive $m_F=0$ state.
\end{thebibliography}
\end{document}